\documentclass[letter,usenatbib, longauth]{mnras}
\usepackage[T1]{fontenc}
\usepackage[utf8]{inputenc}
\usepackage{newtxtext,newtxmath} % MNRAS recommended fonts
\usepackage{graphicx}
\usepackage{amsmath}
\usepackage{newtxtext,newtxmath}
\usepackage{upgreek}
\usepackage{float}
\usepackage{xcolor}
\usepackage{verbatim}
\usepackage{stfloats}
\usepackage[english]{babel}
\usepackage[switch]{lineno}
\newcommand{\kms}{$\rm{\,km \,s}^{-1}$}

\title[JWST reveals CR chemistry in IRAS 07251-0248]{JWST reveals cosmic ray dominated chemistry in the local \\ ULIRG IRAS 07251$-$0248}

\author[G. Speranza et al.]{
G. Speranza$^{1}$\thanks{E-mail: giovanna.speranza@iff.csic.es},
M. Pereira-Santaella$^{1}$,
M. Ag\'undez$^{1}$,
E. Gonz\'alez-Alfonso$^{2}$,
I. Garc\'ia-Bernete$^{3}$,
\newauthor
J. R. Goicoechea$^{1}$,
M. Imanishi$^{4}$,
D. Rigopoulou$^{5,6}$,
M. G. Santa-Maria$^{7,1}$,
N. Thatte$^{5}$
\\
$^{1}$Instituto de F\'isica Fundamental, CSIC, Calle Serrano 123, 28006 Madrid, Spain \\
$^{2}$Universidad de Alcal\'a, Departamento de F\'isica y Matem\'aticas, Campus Universitario, 28871 Alcal\'a de Henares, Spain \\
$^{3}$Centro de Astrobiolog\'ia (CAB), CSIC-INTA, Camino Bajo del Castillo s/n, 28692 Villanueva de la Ca\~nada, Spain \\
$^{4}$National Astronomical Observatory of Japan, 2-21-1 Osawa, Mitaka, Tokyo 181-8588, Japan \\
$^{5}$Department of Physics, University of Oxford, Keble Road, Oxford OX1 3RH, UK \\
$^{6}$School of Sciences, European University Cyprus, Diogenes street, Engomi, 1516 Nicosia, Cyprus \\
$^{7}$Department of Astronomy, University of Florida, P.O. Box 112055, Gainesville, FL 32611, USA \\
}

\date{Accepted 2025 July 17. Received 2025 July 16; in original form 2025 May 6}
\pubyear{\the\year{}}
\begin{document}

\maketitle
%\linenumbers
\begin{abstract}
We analyse the ro-vibrational absorption bands of various molecular cations (HCO$^+$, HCNH$^+$, and N$_2$H$^+$) and neutral species (HCN, HNC, and HC$_3$N) detected in the  \textit{James Webb Space Telescope}/Mid-Infrared Instrument Medium Resolution Spectrometer
spectrum (4.9--27.9\,$\upmu$m) of the local ultra luminous infrared galaxy IRAS~07251$-$0248. 
We find that the molecular absorptions are blueshifted by 160\,km\,s$^{-1}$ relative to the systemic velocity of the target. Using local thermal equilibrium (LTE)  excitation models, we derive rotational temperatures ($T_{\rm rot}$) from 42 to 185\,K for these absorption bands.
This range of measured $T_{\rm rot}$ can be explained by infrared (IR) radiative pumping as a by--product of the strength,  effective critical density, and opacity of each molecular band.
Thus, these results suggest that these absorptions originate in a warm expanding gas shell  ($\dot{M}$$\sim$90--330\,$M_\odot$\,yr$^{-1}$), which might be the base of the larger scale cold molecular outflow detected in this source.
Finally, the elevated abundance of molecular cations 
can be explained by a high cosmic ray ionization rate,  with log($\zeta_{\text{H}_2}$/n$_{\rm H}\, [\text{cm}^3\, \text{s}^{-1}])$
in the range of $-$18.2 (from H$_3^+$) to $-$19.1 (inferred from HCO$^+$ and N$_2$H$^+$, which are likely tracing denser gas),  
consistent with a cosmic ray dominated chemistry as predicted by chemical models. 
\end{abstract}


\section{Introduction}
\label{intro}

Local ultraluminous infrared galaxies (ULIRGs; $L_\text{IR, 8-1000 µm}$ > 10$^{12}$\,$L_{\odot}$) represent a crucial phase in galaxy evolution, as most of them are gas-rich major mergers (e.g., \citealt{Hung14}).  Their nuclei are deeply embedded in dust-obscured environments, with column densities exceeding $N_{\text{H}}$>10$^{24}$\,\text{cm}$^{-2}$ (e.g., \citealt{Alfonso15, Falstad21, Bernete22a, Donnan23}). In such extreme conditions, determining whether the gas primarily fuels active galactic nuclei (AGN), intense starbursts (SB), or both remains challenging. 
To disentangle the dominant excitation mechanisms within ULIRG cores,  interstellar molecular gas has been extensively analysed
(e.g., \citealt{Pereira21, Esposito22, Imanishi2023, Holden2024}). 
These previous works were mostly based on the observation of molecular rotational transitions of CO, HCN, and HCO$^+$ in the mm and sub-mm ranges using the Atacama Large Millimeter/submillimeter Array (ALMA) and the Northern Extended Millimetre Array (NOEMA).
An alternative view of the molecular content is provided by the ro-vibrational molecular bands in the mid-IR. The strongest molecular bands (CO, HCN, and C$_2$H$_2$) have been detected in some  galaxies using space (\textit{Spitzer} and \textit{AKARI}) and ground-based observations (e.g., \citealt{Spoon2004,Lahuis2007,Baba2018,Onishi2021}). Recently, thanks to its superb sensitivity and spatial and spectral resolutions, \textit{James Webb Space Telescope} (JWST) has opened a new window for studying the rich chemistry of ULIRGs through molecular bands from less abundant species (e.g., \citealt{Alfonso2024, Pereira24b, Bernete24a, Buiten24, Buiten24b}).

In this Letter, we analyse the absorption bands of molecular cations in the rich mid-IR spectrum of the eastern nucleus of the ULIRG IRAS~07251$-$0248 (hereafter IRAS~07251; $d$=400\,Mpc;  z = 0.0878; $L_{\text{IR}}$=12.45\,$L_{\odot}$; \citealt{Pereira21}). This nucleus dominates the $L_{\rm IR}$ of the system and host a deeply dust embedded compact core as indicated by the deep 9.7\,$\upmu$m silicate absorption, the strong OH absorptions in the far-IR \citep{Alfonso2017}, and a blueshifted CO $J$=2$-$1 absorption toward the nuclear 230\,GHz continuum \citep{Lamperti22}.

IRAS~07251 also has one of the highest H$_3^+$ abundances, derived from the ro-vibrational $v_2$ band of H$_3^+$ in a sample of local ULIRGs,  as reported by \citet{Pereira24} using JWST\slash NIRSpec observations.
 Specifically these authors measured $N$(H$_3^+$)\slash $N_{\rm H}$ = 4.9 $\times \, 10^{-7}$ for this target, in which the H$_3^+$ $\nu_2$ band is detected in absorption.
H$_3^+$ is a key molecule in cosmic-ray dominated regions (CRDR), the expected environment for ULIRGs, where UV and X-ray photons are mostly absorbed (e.g., \citealt{Papadopoulos10}).  We note that, considering the difficulties in detecting H$_3^+$ before the advent of JWST, alternative diagnostics have been studied to estimate the cosmic ray ionization rate (CRIR; see \citealt{Fontani17, Bovino20, Redaelli24}). Observations and theoretical models suggest that CRIR are attenuated when entering dense clouds (e.g., \citealt{Padovani13, Socci24}), but enhanced by the action of shocks, as in protostellar-like environments (e.g., \citealt{Gaches18, Pineda24}).

 Additionally, H$_3^+$ initiates the chain of chemical reactions that enrich the interstellar medium (ISM; e.g., \citealt{Agundez13, Oka13}).  
  Therefore, molecular cations are expected to be abundant in CRDRs. However, 
with the notable exception of HCO$^+$, observations of molecular cations remain primarily limited to Galactic studies. HCO$^+$ and N$_2$H$^+$ have been extensively studied within the Milky Way (e.g., \citealt{Pety17, Fuente19, Agundez19, Santa21}). 
 \citet{Podio14} and \citet{Ceccarelli14} suggested that the HCO$^+$/N$_2$H$^+$ abundance ratio decreases with increasing cosmic ray (CR) ionization rate, but the detection of both cations is reported in only a few extragalactic studies (e.g., \citealt{Meier12, Donaire23}). 
Similarly, HCNH$^+$, which forms via protonation of HCN or HNC, has been observed in only one extragalactic case (\citealt{Harada24}).  IRAS~07251 therefore represents an ideal target for detecting and characterizing such molecular cations, which remain almost unexplored in extragalactic environments.

We present an analysis of the mid-infrared spectrum of IRAS~07251,
observed with the Mid-Infrared Instrument (MIRI; \citealt{Glasse15}) using the Medium Resolution Spectrometer (MRS; \citealt{Wells15}) onboard JWST. MIRI MRS covers the 4.9--27.9\,$\upmu$m range with a resolving power of R$\sim$1300--3700 (\citealt{Labiano21, Argyriou23}). 
This study represents the first attempt to characterize ISM chemistry in an extragalactic source using this molecular suite.

\begin{table}
\caption{Molecular absorption bands}
\label{tab:molecules}
\centering
\setlength{\tabcolsep}{4.5pt} % tighter column spacing
\footnotesize
\begin{tabular}{lccccc}
\hline \hline
Name\textsuperscript{(1)} & $\lambda$ \textsuperscript{(2)} & $T_{\text{rot}}$ \textsuperscript{(3)} & $N$ \textsuperscript{(4)} & $N$/$N_{\rm H}$ \textsuperscript{(5)} & $f$ \textsuperscript{(6)} \\
 & [$\upmu$m] & [K] & [10$^{15}$ cm$^{-2}$] & [10$^{-8}$] & \\
\hline
HCO$^{+}$ $v_2$        & 12.07 & \,\,42$\,\pm\,14$     & 33.9$\,\pm\,3.1$   & 18   & 0.7 \\
HCNH$^{+}$ $v_4$       & 12.49 & 172$\,\pm\,24$        & 16.3$\,\pm\,1.3$   & 8.6  & 0.7 \\
N$_2$H$^{+}$ $v_2$     & 14.59 & \,\,48$\,\pm\,30$     & \,\,2.1$\,\pm\,0.5$ & 1.1 & 0.7 \\
HC$_3$N $v_5$          & 15.08 & 122$\,\pm\,60$        & 10.0$\,\pm\,0.7$   & 5.3  & 0.7 \\
HCNH$^{+}$ $v_5$       & 15.51 & 152$\,\pm\,19$        & \,\,3.6$\,\pm\,0.2$ & 1.9 & 0.7 \\
HNC $v_2$              & 21.61 & \,\,48$\,\pm\,5$      & 55.7$\,\pm\,7.4$   & 29   & 0.14$\,\pm\,0.08$ \\
\hline
HCN$^\dagger$ $2v_2$   & 7.10  & 185$\,\pm\,25$        & 2000$\,\pm\,600$   & 1100 & 0.6 \\
HCN$^\dagger$ $v_2$    & 14.04 & 100$\,\pm\,25$        & 660$\,\pm\,240$    & 350  & 0.7 \\
\hline
\end{tabular}

\vspace{3pt}
\begin{minipage}{1.\linewidth}
\footnotesize
\textbf{Notes.} 
\textsuperscript{(1)} Ro-vibrational band; 
\textsuperscript{(2)} Rest-frame wavelength of the Q(1) transition, except for HCN $2\nu_2$, for which the R(1) transition is reported; 
\textsuperscript{(3)} Rotational temperature; 
\textsuperscript{(4)} Column density; 
\textsuperscript{(5)} Fractional abundance adopting $N_{\rm H} = 1.89 \times 10^{23}$ cm$^{-2}$ from \citet{Pereira24}; 
\textsuperscript{(6)} Covering factor.\\
$^\dagger$ Values for HCN are from \textcolor{blue}{García-Bernete et al., submitted}.
\end{minipage}
\end{table}

\section{Analysis and results}
\label{analysis}

\begin{figure*}
\centering
\includegraphics[width=0.49\textwidth]{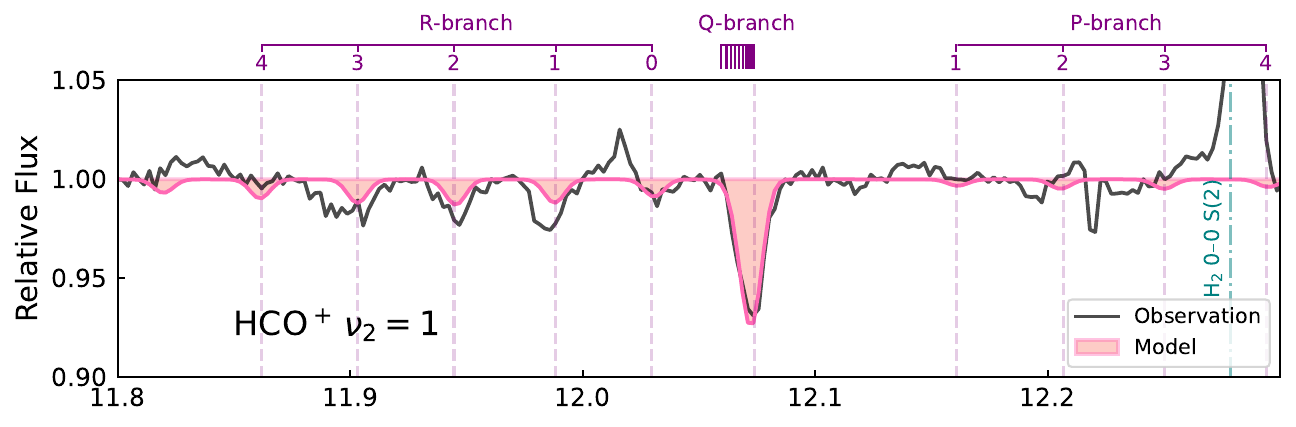}
\includegraphics[width=0.48\textwidth]{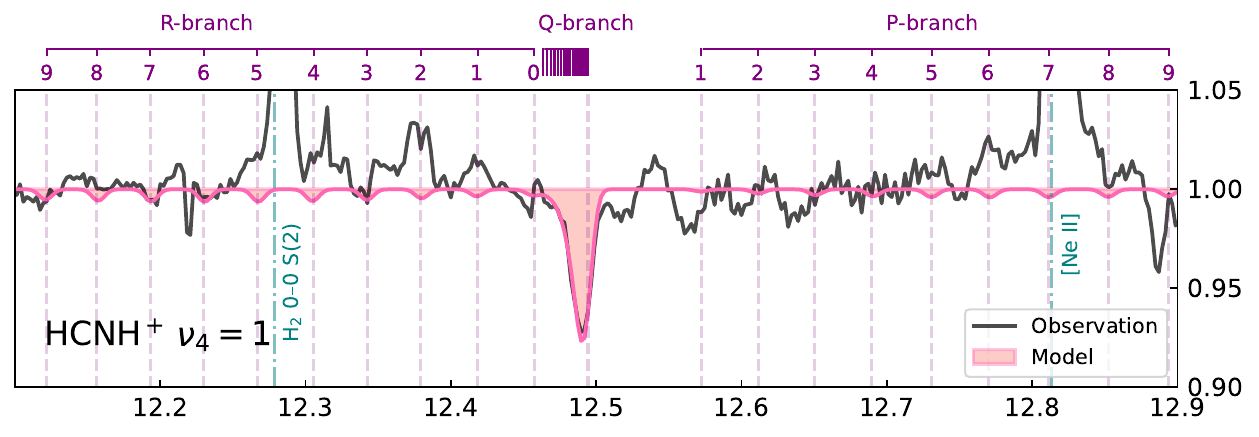}
\includegraphics[width=0.49\textwidth]{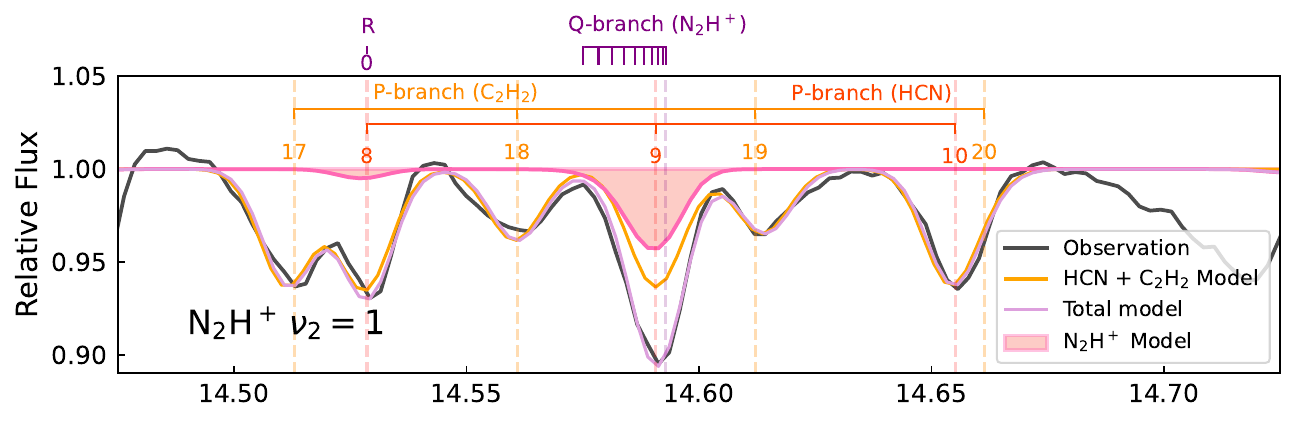}
\includegraphics[width=0.48\textwidth]{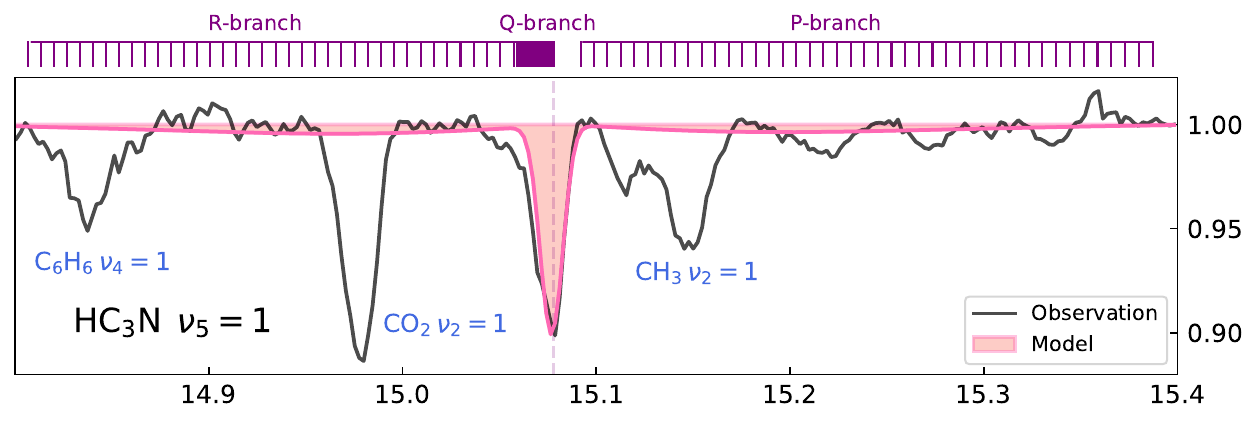}
\includegraphics[width=0.495\textwidth]{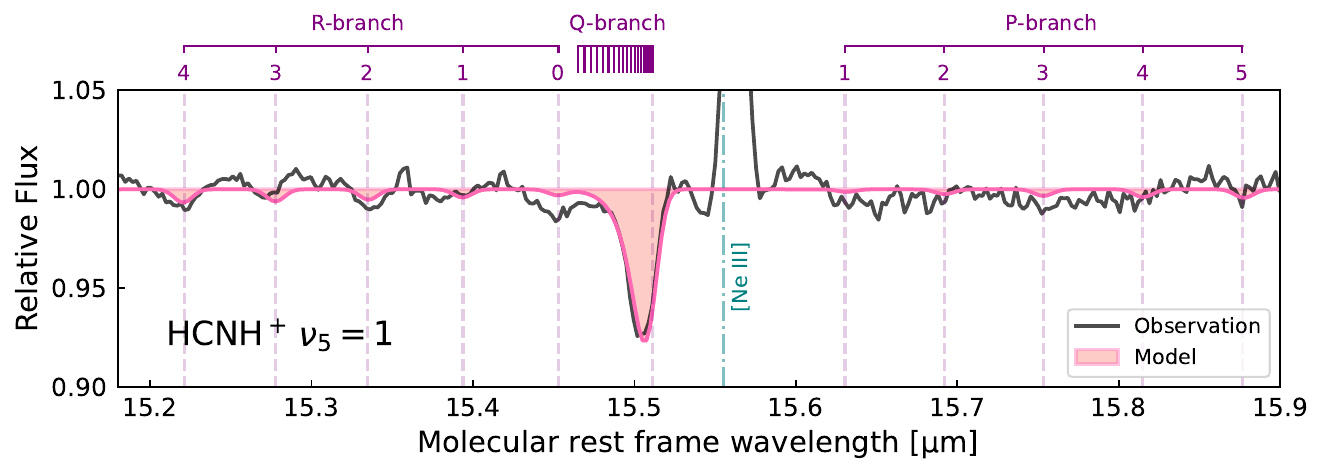}
\includegraphics[width=0.47\textwidth]{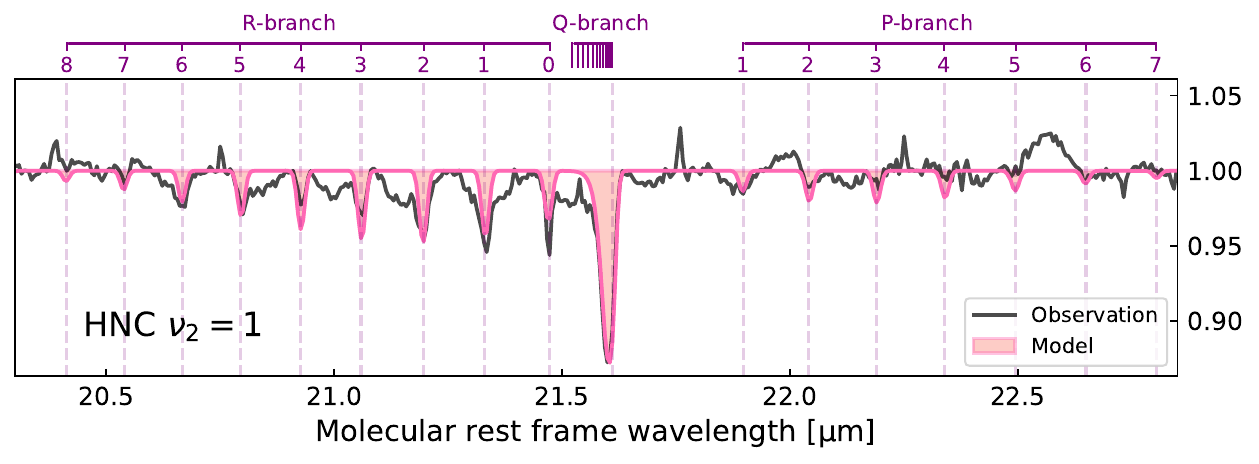}
\caption {LTE models and continuum-normalized observed spectra of the fundamental ro-vibrational bands detected in absorption:  from left to right and top to bottom HCO$^+$ $v_2$, HCNH$^+$ $v_4$, N$_2$H$^+$ $v_2$,  HC$_3$N $v_5$, HCNH$^+$ $v_5$, and HNC $v_2$. The observed spectra are shown in black, models are represented by pink line and shaded area. Vertical dashed purple lines indicate the wavelength of the resolved transitions of the P- and R-branches and the origin of the Q-branch. The numbers at the top of each panel indicate the rotational number $J$ of the lower level of the transition.
 All panels are displayed in the molecular rest frame, which is blueshifted by $\sim 160$ km s$^{-1}$ relative to the systemic velocity of this target (based on CO $J$=2--1; \citealt{Pereira21}).
The wavelengths (in the molecular rest frame) of the detected emission lines are marked by dashed-dotted green lines (H$_2$ 0-0 S(2)  at 12.28 $\mu$m, [Ne~II]$\lambda$12.81 $\mu$m, and [Ne~III]$\lambda$15.56 $\mu$m). The N$_2$H$^+$ panel includes the model of the  P(17) to P(20) transitions of C$_2$H$_2$ $v_5$ and the  P(8) to  P(10) transitions of HCN $v_2$ (dashed orange vertical lines and the orange model). The total model is shown in violet, while the N$_2$H$^+$ model is displayed in pink with a shaded area, as the rest of detections.}
\label{fig:molecules}
\end{figure*}

The mid-IR spectrum of the eastern nucleus of IRAS 07251 shows absorptions associated with several ro-vibrational bands of neutral molecules and cations. In particular, we investigate the fundamental bands of HCO$^+$ $v_2$, HCNH$^+$ $v_4$ and $v_5$, N$_2$H$^+$ $v_2$, HC$_3$N $v_5$, and HNC $v_2$ (see Table~\ref{tab:molecules}), complemented by the HCN 2$v_2$ and $v_2$ bands analysed by \textcolor{blue}{García-Bernete et al. submitted}. 
Details on data reduction and the molecular spectroscopic parameters are provided in Appendices~\ref{AppendixA} and \ref{apx:spec_param}, respectively.
All measurements are reported in Table~\ref{tab:molecules} and shown in Fig.~\ref{fig:molecules}.

We estimated the rotational temperature ($T_{\text{rot}}$) and column density (Table~\ref{tab:molecules}) of each vibrational band by fitting the spectra using a local thermal equilibrium (LTE) model.
To do so, we first normalized the observed spectra by the local continuum level, estimated through a spline-interpolated baseline. 
The radiative transfer LTE models assume a background emission source and a homogeneous layer of molecular material in front, whose rotational levels are populated according to a single $T_{\text{rot}}$.
For each molecule, we constructed a grid of 50$\times$50 models where the column density, $N$, and $T_{\rm rot}$ varied.
We find that the molecular absorptions are blueshifted by $-160$\,km\,s$^{-1}$ relative to the restframe velocity of this object (based on CO $J$=2--1; \citealt{Pereira21}). This suggests that these molecular absorptions are produced in an expanding shell and may represent the innermost part of the molecular outflow reported by \citet{Lamperti22}.
We adopted an intrinsic velocity dispersion, $\sigma$, of 105\,\kms\, based on the width of non blended molecular lines (see \textcolor{blue}{García-Bernete et al. submitted}).

We select as best fit model that with the minimum $\chi^2$ between observation and model. 
For all bands except HNC $v_2$, the $\chi^2$ computation was performed only on the spectral range covered by the Q-branch, since the R-branch is only tentatively detected in most bands, and the P-branch is fainter.  The line overlap is taken into account in the fitting procedure.
Although the individual ro-vibrational transition lines are spectrally unresolved in the Q-branch, the fit is sensitive to the rotational temperature since the Q-branch broadens with increasing rotational temperature as higher $J$ levels are populated.
 Before the $\chi^2$ evaluation, a covering factor ($f$) of 0.7  (based on \textcolor{blue}{García-Bernete et al. submitted}) is applied to all models. The covering factor is defined as the ratio between the absorbed continuum and the total continuum (absorbed plus unabsorbed), which is equivalent to the "background factor" defined in \citet{Buiten24b}.
In order to model the Q-branch of N$_2$H$^+$ $v_2$, we accounted for contamination from the HCN $v_2$  P(9) line. To do so, we fitted the HCN $v_2$  P(8) and P(10) lines and interpolated the strength of the  P(9) line from the $T_{\rm rot}$ determined from the other two P transitions. The model for N$_2$H$^+$ $v_2$ is then obtained after removing such contribution. For the HNC $v_2$ band, both the R and the P branches are clearly detected up to $J_{\rm low}=8$. Therefore, we computed $\chi^2$ over the full wavelength range (i.e., 20.30 -- 22.85 $\upmu$m), which includes the R- and P-branches.
In this case, $f$ was treated as free parameter of the model, as the contribution from the colder dust continuum becomes more significant at longer wavelengths (e.g., \citealt{Donnan24}), leading to a lower fraction of absorbed warm continuum.
Indeed, we obtained $f$ = 0.14$\,\pm\,0.08$ for HNC $v_2$ (Table~\ref{tab:molecules}).   

The parameters of the best-fitting models are listed in Table~\ref{tab:molecules} and shown in Fig.~\ref{fig:molecules}.  Errors on the measured $T_{\rm rot}$ and N are calculated from a series of 100 Monte Carlo simulations.  In the case of N$_2$H$^+$, both the total profile and the HCN model have been perturbed to account for the uncertainty due to the line blending.
The resulting $T_{\rm rot}$ range from 40\,K to 185\,K. The HCNH$^+$ $v_4$ and $v_5$ and HCN 2$\nu_2$ bands reach the highest temperatures ($\sim 150-185$\,K), while the rest of cations and neutral molecules exhibit lower $T_{\rm rot}$ between 40 and 120\,K.

\section{Discussion}
\label{discussion}

\subsection{IR radiative pumping to explain the observed $T_{\rm rot}$ range} 
\label{lvg}

As shown in Table~\ref{tab:molecules}, $T_{\rm rot}$ vary by up to $\sim$140 K ranging from the lowest temperature measured for HCO$^+\,v_2$ (42 K) to the highest reported for HCNH$^+\,v_4$ (172 K)  and HCN 2$\nu_2$ (185 K). 
Based on the analysis of the strong molecular bands (HCN, C$_2$H$_2$, and H$_2$O) in this source by \textcolor{blue}{Gonz\'alez-Alfonso et al. in prep.}, we expect all these bands to originate in a relatively small volume ($r$< 20-75\,pc) close to the compact IR core of IRAS~07251. Therefore, in this section, we investigate if radiative pumping excitation can explain the wide range of observed $T_{\rm rot}$.

Different $T_{\rm rot}$ are expected due to the different responses of each molecule when exposed to an IR radiation field (Einstein $B$ coefficients) and their different collisional excitation  rate coefficients ( thus, different critical densities).
These effects were discussed by \citet{Imanishi16} to explain the ground state and vibrationally excited HCN\slash HCO$^+$\slash HNC $J=3-2$ emissions in the ULIRG IRAS 20551$-$4250, where IR radiative pumping is the dominant excitation mechanism.

Here, we further explore whether IR radiation can produce the observed $T_{\text{rot}}$ for these bands by performing 
nonlocal thermodynamic equilibrium (NLTE) models under the large velocity gradient (LVG) formalism. For these NLTE models, we assume
an illuminating blackbody with temperatures ($T_{\text{rad}}$) from 200 to 500 K with a dilution equivalent to  a cloud located at a distance of 50\,pc from  a source with a radius of 20\,pc, a kinetic temperature ($T_{\text{kin}}$) of $\sim$200\,K ( based on measurements from  \textcolor{blue}{Gonz\'alez-Alfonso et al. in prep.} and \textcolor{blue}{García-Bernete et al. submitted}), and $n_{\text{H}_2} = 10^{4} \text{cm}^{-3}$ as fiducial values.
Then, we determined the apparent $T_{\rm rot}$ of the simulated NLTE absorption band by finding the best fit model from the LTE grids used in Sect.~\ref{analysis}.
 
The left panel of Fig.~\ref{fig:Trad} shows the apparent $T_{\rm rot}$ of the NLTE models as a function of $T_{\rm rad}$. 
The different response of each molecule to the radiation field is clear in this figure.
The HCNH$^+$ curve is the closest to $T_{\text{rot}}$ = $T_{\text{rad}}$ at  all $T_{\text{rad}}$, whereas HCO$^+$ is the least efficient reacting to the radiation field.

For HCO$^+\, v_2$, its lower trend with respect to the other curves is due to its relatively low Einstein $B$ coefficient, which is the smallest among the studied bands (2 to 13 times lower than the others),  excluding HCN 2$\nu_2$, whose high column density significantly affects the observed $T_{\rm rot}$, as explained later in this Section (see also Table~\ref{tab:coeff}).
Conversely, the behavior of the HCNH$^+$ curve is  influenced by the much lower dipole moment (about 10 times below the other molecules, also reported in Table~\ref{tab:coeff}), leading to a lower  effective critical density  (as shown in the right panel of Fig.~\ref{fig:Trad}), thus to a more efficient excitation of higher J  levels, and consequently to a broadening of the Q branch.

As discussed above, the width of the Q-branch absorption traces $T_{\text{rot}}$ of the ground state. For increasing $n_{\text{H}_2}$, $T_{\text{rot}}$ increases, tending to $T_{\rm kin}$, and making broader the observed Q-branch absorption. Therefore, we define the  effective critical density, $n_{\rm crit}^{\rm eff}$, for each band as the value of $n_{\text{H}_2}$ at which the observed full width at half-maximum (FWHM) of the Q-branch reaches 99$\%$ of the FWHM of the band under LTE conditions (i.e., $T_{\text{rot}}$=$T_{\rm kin}$).
We note that we cannot adopt the critical densities of the individual rotational transitions (e.g., \citealt{Shirley15}) since the Q-branch is sensitive to the population of all the ground state rotational levels.
We computed NLTE models with $T_{\rm kin}$ from 50 to 200 K and $n_{\text{H}_2}$ ranging from 10$^4$ to 10$^{9}$ cm$^{-3}$, with the same column densities reported in Table~\ref{tab:molecules}. 
The right panel of Fig.~\ref{fig:Trad} shows $n_{\rm crit}^{\rm eff}$ of each band as a function of $T_{\text{kin}}$. The HCNH$^+$ curve stands apart from the others, with $n_{\rm crit}^{\rm eff}$ more than one dex lower than those of the other bands. This indicates that collisions with H$_2$ play a significant role in allowing HCNH$^+$ to reach equilibrium more efficiently with radiation since $T_{\text{rot}}$ starts at higher values than for the other molecules (see the left panel of Fig.~\ref{fig:Trad}). 
 
 The opacity of the band can also significantly affect the observed $T_{\rm rot}$. This is illustrated by the two HCN curves in Fig.~\ref{fig:Trad}, which correspond to the range of HCN column density values from $N$=2.0$\times \,10^{18}$\,cm$^{-2}$ (upper curve) to $N$=6.6$\times\,10^{17}$ cm$^{-2}$ (lower curve; 
 see Table~\ref{tab:molecules}).
Thus, higher column densities, hence higher opacities, can increase the observed $T_{\rm rot}$ of the bands.

Consequently, if IR pumping, NLTE excitation, and opacity effects are taken into account, it is plausible that all these molecular bands originate at the same physical region close to the core of IRAS~07251, even if they exhibit different observed $T_{\rm rot}$ (Table~\ref{tab:molecules}).

\begin{figure*}%[t]
\centering
\includegraphics[width=1\textwidth]{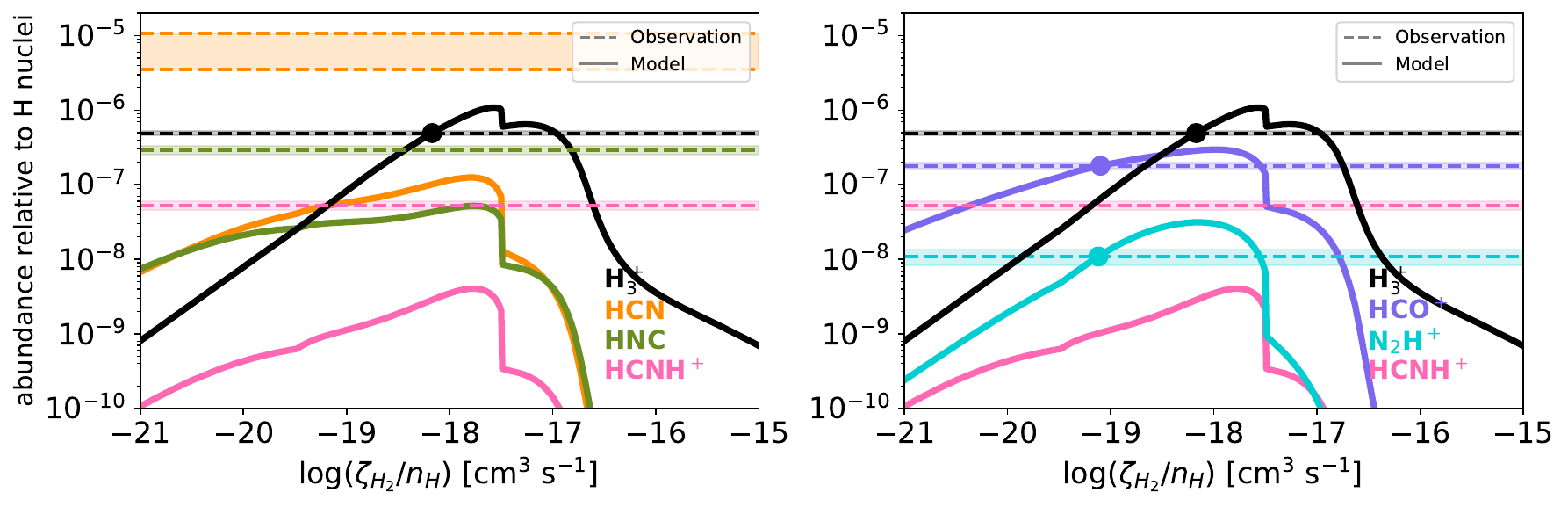}
\caption {Predicted fractional abundances by chemical models as function of the $\zeta_{\text{H}_2}/n_{\text{H}_2}$ ratio (solid curves) and observed values (horizontal dashed lines) with shaded area for the corresponding uncertainties.
We assumed a H column density of $N_{\text{H}} = 1.89 \times 10^{23} \text{cm}^{-2}$ according to \citet{Pereira24}.
H$_3^+$ (black), HCN (orange), HNC (green), HCNH$^+$ (pink) are presented in the left panel. Both observed abundances from \textcolor{blue}{García-Bernete et al. submitted} are shown for HCN, with the uncertainty represented by their difference. The median value is shown for the observed HCNH$^+$ abundances reported in Table~\ref{tab:molecules}.
The cations H$_3^+$ (black), HCO$^+$ (purple), N$_2$H$^+$ (turquoise), and HCNH$^+$ (pink) are shown in the right panel.
The black point indicates where the H$^+_3$ model coincides with the observed abundance from \citet{Pereira24},  while the purple and turquoise points correspond to the intersections between observations and models of HCO$^+$ and N$_2$H$^+$,  respectively.}
\label{fig:models}
\end{figure*}
\subsection{Comparison with chemical models} \label{ss:chemical}

To shed light on the origin of the observed molecules, we carried out chemical modelling calculations using a 
model for obscured dense molecular clouds \citep{Agundez13}. We adopted a gas $T_{\rm kin}$ of 200\,K  (according to the gas temperature reported by \textcolor{blue}{García-Bernete et al. submitted}), 
a high visual extinction of 30 mag to prevent UV photons from playing a role, roughly solar elemental abundances including dust depletion (table 3 of \citealt{Agundez13}; see also Appendix~\ref{apx:chemical} for a discussion on this assumption), and the gas-phase chemical network corresponding to the latest release of the UMIST Database for Astrochemistry \citep{Millar24}. We varied the cosmic-ray ionization rate of H$_2$ to cover the range $\zeta_{\text{H}_2}$/$n_{\text{H}}$=10$^{-21}$-10$^{-15}$\,cm$^{3}$\,s$^{-1}$. We note that,  testing the results through variations in $n_{\rm H}$, steady state abundances, in which we focus here, are only sensitive to the $\zeta_{\text{H}_2}$/$n_{\text{H}}$ ratio, but not to these two parameters separately  (see e.g., \citealt{Ceccarelli14, Padovani09}). {In Fig.~\ref{fig:models} we compare the calculated abundances relative to H nuclei  with the observed values as a function of the $\zeta_{\text{H}_2}$/$n_{\text{H}}$ ratio}. We observe that the H$_3^+$ prediction matches the corresponding observed value at 
a log($\zeta_{\text{H}_2}$/n$_{\rm H}\, [\text{cm}^3\, \text{s}^{-1}]) $ of $\sim$ -18.2.

The left panel of Fig.~\ref{fig:models} shows H$_3^+$ together with HCN, HNC, and HCNH$^+$, which are closely related through the protonation reactions  H$_3^+$+HCN(HNC) $\to$ HCNH$^+$ + H$_2$, the dissociative recombination reaction HCNH$^+$ + e$^-\to$ HCN(HNC) + H, as well as the neutral reaction HNC + H $\to$ HCN + H.
The ion route involving HCNH$^+$ is expected to produce HCN and HNC with similar yields. However, the neutral-neutral reaction can isomerize HNC into HCN at high temperatures
(\citealt{Schilke92}), as its rate coefficient k(T) = 10$^{-10} e^{\text{E}_a/\text{T}}$ depends on the activation energy barrier ($\text{E}_a$), which in our models is set at 1200\,K, following chemical models of ISM clouds 
(\citealt{Graninger14}). 
We observe that models underestimate the absolute abundances of these species by a factor of $\sim$10--100. 
However, the relative abundance HCNH$^+$/(HCN + HNC) ratio is consistent between models and observations within a factor of $<$4 (0.019 vs. 0.005--0.014) since this ratio is driven by the protonation reactions and mostly depends on the $\zeta_{\text{H}_2}$/n$_{\text{H}}$ ratio and their relative abundances.
However, the predicted HCN/HNC abundance ratio is lower than the observed ratio (2.3 vs. 12--37). This suggests that the neutral reaction has not begun depleting HNC and might indicate that the barrier of this reaction is lower than the assumed value (e.g., \citealt{Hirota98}) or that the H abundance is higher than that predicted by the models, which depends on the relatively uncertain assumed H$_2$ formation rate (3$\times$10$^{-17}$\,cm$^{3}$\,s$^{-1}$; \citealt{Wakelam2017}).

In the right panel of Fig.~\ref{fig:models}, we present the fractional abundances predicted for all the detected cations (Table.~\ref{tab:molecules}). As expected from previous studies (e.g., \citealt{Agundez18, Gaches19}), HCO$^+$ is the most abundant cation, after H$_3^+$, in both models and observations. There is a good agreement, within a factor of 4, for HCO$^+$ and N$_2$H$^+$. 
However, HCNH$^+$ is underestimated by the models.
This discrepancy likely arises from the underestimation of HCN and HNC (as discussed above), which leads to lower predicted abundances of HCNH$^+$. 
 Observations of HCO$^+$ and N$_2$H$^+$ intersect the models at log($\zeta_{\text{H}_2}$/n$_{\rm H}\, [\text{cm}^3\, \text{s}^{-1}]) \sim$ $-$19.1, differently from the H$_3^+$ intersection at log($\zeta_{\text{H}_2}$/n$_{\rm H}\, [\text{cm}^3\, \text{s}^{-1}]) \sim$ $-$18.2 \footnote{There is another solution at $\log \zeta_{\text{H}_2}$/n$_{\rm H}\sim-$17.5, however this solution lies close to a discontinuity in the molecular abundances, which is related to the bistability of dark cloud chemistry \citep{LeBourlot1993}. Near this discontinuity, a slight increase of $\zeta_{\text{H}2}/n_{\rm H}$ results in a sharp decrease in the molecular abundances. Thus, although this solution is physically possible, we favor the lower $\zeta_{\text{H}_2}$/n$_{\rm H}$ solution, which lies in a more stable regime of the chemical models.} This discrepancy may arise because the model adopts a single constant density for all molecules, suggesting that H$_3^+$ may have a slightly lower density compared to HCO$^+$ and N$_2$H$^+$.  This is consistent with predictions from theoretical models (e.g., \citealt{Padovani11, Padovani13}) and  observational evidence (e.g., \citealt{Sabatini23, Socci24}), which suggest that ionization from cosmic rays depends on local physical conditions and may be attenuated when penetrating denser gas regions. In conclusion, our results support a scenario in which the observed molecules originate from a CRDR, with H$_3^+$ playing a key role in shaping the chemistry of the ISM (see also models from \citealt{Harada10} and \citealt{Bayet11}).

\section{Summary and conclusions}  
\label{conclusions}

We analysed absorption features from the fundamental ro-vibrational bands of HCO$^+$ $\nu_2$, HCNH$^+$ $\nu_4$ and $\nu_5$, N$_2$H$^+$ $\nu_2$, HC$_3$N $\nu_5$, and HNC $\nu_2$  detected by JWST/MIRI MRS in the eastern nucleus of the ULIRG IRAS 07251$-$0248 ($d$$\sim$400 Mpc). Based on the physical properties inferred from these bands and the comparison with chemical models, we draw the following conclusions:

\begin{itemize}

\item \textit{Outflowing warm molecular gas.} We find that the molecular absorptions are blueshifted by 160\,km\,s$^{-1}$ relative to the rest-frame velocity of this object.  From the LTE models, we find high rotational temperatures: HCNH$^+$ $\nu_4$ and $\nu_5$ and HCN 2$\nu_2$ exhibit the highest $T_{\rm rot}$ from $\sim$150 to $\sim$185 K, while 40--120\,K, are measured for the remaining species. This suggests that the absorptions are produced in a expanding warm molecular shell. Using equation 11 of \citet{Alfonso2017} and assuming a shell radius of 20--75\,pc, the observed $N_{\rm H}$ and velocity correspond to a mass outflow rate of $\dot{M}$$\sim$90--330\,$M_\odot$\,yr$^{-1}$, and could be the inner launching region of the larger scale (170\,pc) and faster (380\,km\,s$^{-1}$) cold molecular outflow ($\dot{M} \sim 50\,\mathrm{M}_\odot\,\mathrm{yr}^{-1}$) reported by \citet{Lamperti22}.

\item \textit{Evidence for IR radiative pumping}. The observed $T_{\rm rot}$ values vary by up to 140 K. These differences can be explained by a radiative transfer model under NLTE condition. In this context, infrared radiative pumping is a viable excitation mechanism. This supports the idea that all molecules originate from the same shell near the core of IRAS 07251. 

\item \textit{CRDR chemistry in IRAS 07251}. We find that CR play a key role in the chemistry of the nuclear ISM in IRAS 07251. According to chemical models, the fractional abundances of the observed molecular cations (H$_3^+$, HCO$^+$, HCNH$^+$, and N$_2$H$^+$) are consistent with a CRDR with a high log($\zeta_{\text{H}_2}$/n$_{\rm H}\, [\text{cm}^3\, \text{s}^{-1}])$  in the range of -19.1 to -18.2. 
 The lower value is derived from HCO$^+$ and N$_2$H$^+$, which are believed to trace denser gas than H$_3^+$. This result is consistent with theoretical predictions suggesting a decrease in cosmic ray ionization rate in denser environments.

\end{itemize}

This study demonstrates the capability of JWST to probe the ISM chemistry through several molecular ro-vibrational bands, transitions that were previously unexplored in extragalactic sources due to limitations in sensitivity and resolution. These bands are also crucial for probing the excitation mechanisms dominating the region near the galaxy's core.

\section*{Acknowledgements}

The authors acknowledge the GOALS team for developing their observing programs.
GS and MPS acknowledge support under grant CNS2023-145506 funded by MCIN/AEI/10.13039/501100011033 and the European Union NextGenerationEU/PRTR. 
MPS acknowledges support under grant RYC2021-033094-I funded by MCIN/AEI/10.13039/501100011033 the European Union NextGenerationEU/PRTR.
MPS, EGA, JRG and MGSM thank the Spanish MCINN for funding support under grant PID2023-146667NB-I00 funded by MCIN/AEI/10.13039/501100011033.
MA acknowledges support under grant PID2023-147545NB-I00.
IGB is supported by the Programa Atracci\'on de Talento Investigador ``C\'esar Nombela'' via grant 2023-T1/TEC-29030 funded by the Community of Madrid. EG-A thanks the Spanish MICINN for support under project PID2022-137779OB-C41. 
This work is based on observations made with the NASA/ESA/CSA James Webb Space Telescope. The data were obtained from the Mikulski Archive for Space Telescopes at the Space Telescope Science Institute, which is operated by the Association of Universities for Research in Astronomy, Inc., under NASA contract NAS 5-03127 for JWST; and from the European JWST archive (eJWST) operated by the ESAC Science Data Centre (ESDC) of the European Space Agency. These observations are associated with program \#3368.

\section*{Data availability} 
The data are available in the JWST Science Archive at https://jwst.esac.esa.int/archive/, under proposal ID $\#$3368.

\bibliographystyle{mnras}
\bibliography{sample}

\appendix
\section{MIRI\slash MRS data reduction}\label{AppendixA}

IRAS~07251$-$0248 was observed with JWST as part of the Cycle 2 Large Program \#3368 (P.I. L. Armus and A. Evans). In this Letter, we analyse the integral-field spectroscopy obtained with MIRI\slash MRS in the 4.9--28.1\,$\upmu$m range with a spectral resolution of $R\sim$1300--3700 \citep{Labiano21}.

The raw data were processed using the JWST calibration pipeline (version 1.12.4; \citealt{Bushouse_1_12_4}) and context 1253. In addition to the standard data reduction steps, we performed a number of custom steps to reduce the effect of cosmic rays and hot and cold pixels.
We extracted the spectrum of the E nucleus of IRAS~07251$-$0248 assuming that it is dominated by a point source at the angular resolution of MRS (0\farcs25--0\farcs60 depending on the wavelength). Further details on the data reduction and extraction of the nuclear spectrum can be found in \citet{Pereira2022}, \citet{Bernete22b} and \citet{GarciaBernete2024_ice}.

\section{Molecular parameters}\label{apx:spec_param}

We analysed the mid-IR ro-vibrational absorption bands of five molecules (HCO$^+$, HCNH$^+$, N$_2$H$^+$, HC$_3$N, and HNC). In this appendix, we describe the spectroscopic parameters as well as the collisional rate coefficients used in this Letter.

For HNC, we used the line list provided by \citet{Barber2014}, part of the ExoMol database, and the collisional coefficients from \citet{Dumouchel2010} and \citet{HernandezVera2017}. The HITRAN database \citep{Gordon2022} was used for the HC$_3$N line list together with the collisional coefficients calculated by \citet{Faure2016}.

For the remaining three molecular cations (HCO$^+$, HCNH$^+$, and N$_2$H$^+$), we used the wavelengths of rotational and ro-vibrational transitions from laboratory measurements to determine the best fit molecular spectroscopic constants of all the fundamental vibrational states. We performed this fit using PGOPHER \citep{Western2017}.
All these cations are linear molecules with a $^1\Sigma$ ground electronic state. Therefore, within PGOPHER, we used a Hamiltonian with a rotational operator, including up to the sextic ($H$) and quartic ($D$) centrifugal distortion constants for the ground and vibrationally excited states, respectively, and, for the bending modes, we also included the l-type interaction with the l-type doubling constant $q$ and its high order correction, $q_{\rm D}$, as free parameters. The origin of the band is also determined during the fit.

We used the strength of the vibrational bands, calculated theoretically or measured in the laboratory when available, to determine the
Einstein $A$ coefficient of the band, $A_{\rm band}$.
Then, the Einstein $A$ coefficient of each ro-vibrational transition, $A_{ul}$ was computed using the relation $A_{ul}=A_{\rm band}(\nu_{ul}\slash \nu_0)^3  S_{J_l}^{\Delta J} \slash (2J_u + 1)$, where $\nu_{ul}$ and $\nu_0$ are the frequency of the ro-vibrational transition between levels $u$ and $l$ and the origin of the vibrational band, respectively, $J_u$ is the quantum number $J$ of the upper level $u$, and $S_{J_l}^{\Delta J}$ is the H\"onl-London factor (e.g., \citealt{Hansson2005}). The line lists and spectroscopic constants are available from the authors upon request and will be presented in a future work (Pereira-Santaella in prep.).

We note that the fundamental $v_2$ bands of HCO$^+$ and N$_2$H$^+$ are included in the CDMS database \citep{Endres2016}. For these two vibrational bands, we obtain wavelengths and Einstein $A$ coefficients in good agreement with those in CDMS with differences $<$0.5\,km\,s$^{-1}$ for the position of the lines and $<$2\,\% for the $A$ coefficients.

\subsection{HCO$^+$}
For the fit of the spectroscopic constants, we used the rotational transitions presented in \citet{Hirota1988, Lattanzi2007, Hirao2008}, and \citet{Cazzoli2012}, the $v_2$ fundamental band measured by \citet{Davies1984} and \citet{Kawaguchi1985}, and the $v_1$ and $v_3$ stretching modes from \citet{Amano1983, Davies1984b, Foster1984, Neese2013} and \citet{Siller2013}.
The strength of the bands are taken from the experiments by \citet{Keim1990} and the calculations by \citet{Martin1993}.
The collisional rate coefficients with H$_2$ are from \citet{DenisAlpizar2020}.

\subsection{N$_2$H$^+$}

To obtain the spectroscopic constants, we fitted rotational transitions \citep{Verhoeve1990, Amano2005, Cazzoli2012, Yu2015}, fundamental $v_2$ ro-vibrational transitions \citep{Sears1985}, and $v_1$ and $v_3$ transition from \citet{Foster1984b, Nakanaga1990, Sasada1990, Kabbadj1994} and \citet{Kalosi2017}.
The strength of the bands are from the experiments by \citet{Keim1990} and the calculations by \citet{Heninger2003}.
We used the collisional rate coefficients with H$_2$ from \citet{Balanca2020}.

\subsection{HCNH$^+$}

We fitted the rotational transitions \citep{Amano2006, Silva2024}, and the ro-vibrational transitions from the three stretching modes ($v_1$, $v_2$, and $v_3$; \citealt{Altman1984, Kajita1988, Liu1988}) and from the two bending modes ($v_4$ and $v_5$; \citealt{Tanaka1986, Ho1987}) to obtain the spectroscopic constants.
For the strength of the vibrational bands, we calculated the average value from the different levels of quantum theory calculations in the CCCBDB database \citep{Johnson2022}.
The collisional rate coefficients with H$_2$ are from \citet{Bop2023}.

\section{NLTE models}\label{figura}

\begin{table}
\caption{Einstein coefficients, dipole moments, and effective critical densities of the ro-vibrational transitions shown in Fig.\ref{fig:Trad}.}
\label{tab:coeff}
\centering
\setlength{\tabcolsep}{4.5pt} % tighter column spacing
\footnotesize
\begin{tabular}{lccccc}
\hline \hline
Name\textsuperscript{(1)} & A$_{ul}$ \textsuperscript{(2)} & B$_{lu}$/B$_{lu}^{\text{HCO}^+}$ \textsuperscript{(3)} & Dipole &  $n_{\rm crit}^{\rm eff}$ \textsuperscript{(5)} & Ref.\textsuperscript{(6)}\\
& & & moment\textsuperscript{(4)} \\
 & [ s$^{-1}$] &  & [D] &  [cm$^{-3}$] & \\
\hline
HCO$^{+}$ $v_2$     & 1.46   & 1 & 3.89  & 4.33  $\times \, 10^{8}$  & (Y94)  \\
N$_2$H$^{+}$ $v_2$  & 8.13   & 9.8 & 3.4  & 3.80 $\times \, 10^{8}$  & (H90)  \\
HCNH$^{+}$ $v_5$   & 9.14    & 13.3 &  0.29 & 9.71  $\times \, 10^{5}$  &  (B86)\\
HNC $v_2$        & 3.37      & 13.2 &  3.05 & 7.03 $\times \, 10^{7}$  & (B76)\\
HCN $2v_2$   & 2.09  &    0.3  &  2.98 & 7.03  $\times \, 10^{7}$ & (E84) \\
HCN $v_2$ & 1.83   & 2.0 & 2.98  & 7.03  $\times \, 10^{7}$  & (E84) \\
\hline
\end{tabular}

\vspace{3pt}
\begin{minipage}{1.\linewidth}
\footnotesize
\textbf{Notes.} 
\textsuperscript{(1)} Vibrational band; 
\textsuperscript{(2)} Einstein A$_{ul}$ coefficient corresponding to the Q(1) transition, except for HCN $2\nu_2$  in which case it corresponds to the R(1) transition; 
\textsuperscript{(3)} Einstein B$_{lu}$ coefficient normalized to the HCO$^+\, \nu_2$ B$_{lu}$ coefficient; 
\textsuperscript{(4)} Dipole moment;
\textsuperscript{(5)} $n_{\rm crit}^{\rm eff}$  at 200 K; \textsuperscript{(6)} References for the dipole moment: (Y94) \citet{Yamaguchi94}, (H90) \citet{Havenith90}, (B86) \citet{Botschwina86}, (B76) \citet{Blackman76}, (E84) \citet{Ebenstein84}. \\

\end{minipage}
\end{table}

\begin{figure*}
\centering
\includegraphics[width=.49\textwidth]{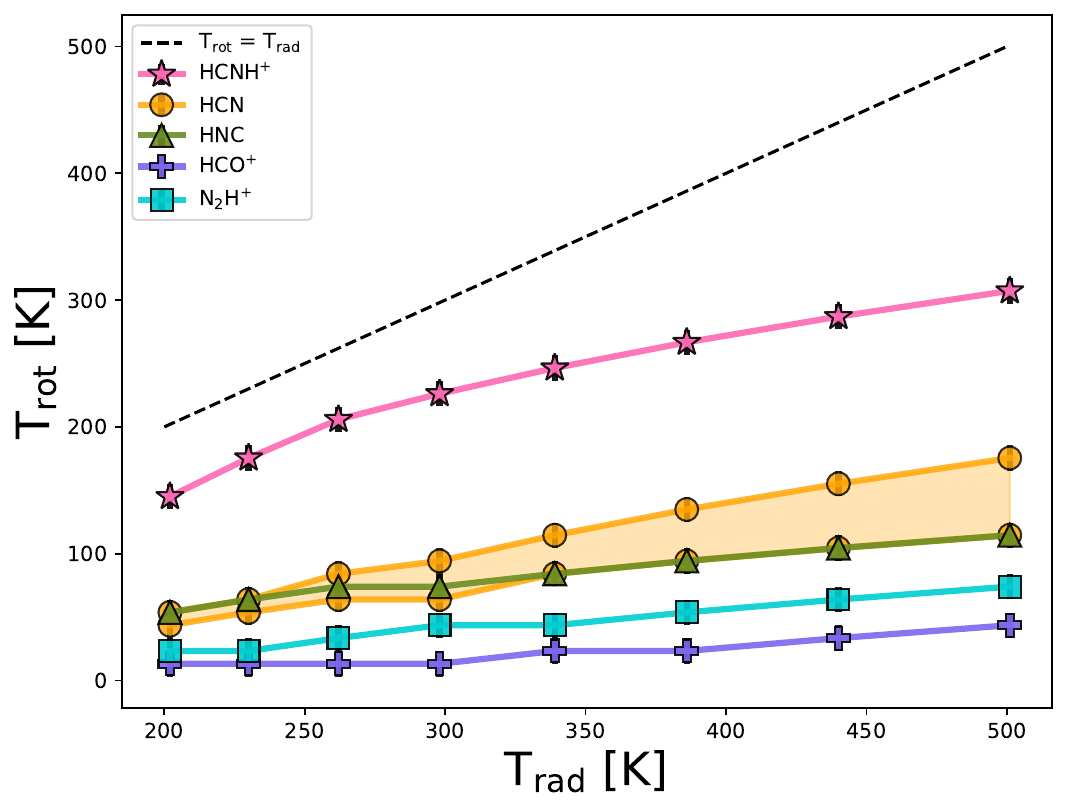}
\includegraphics[width=.49\textwidth]{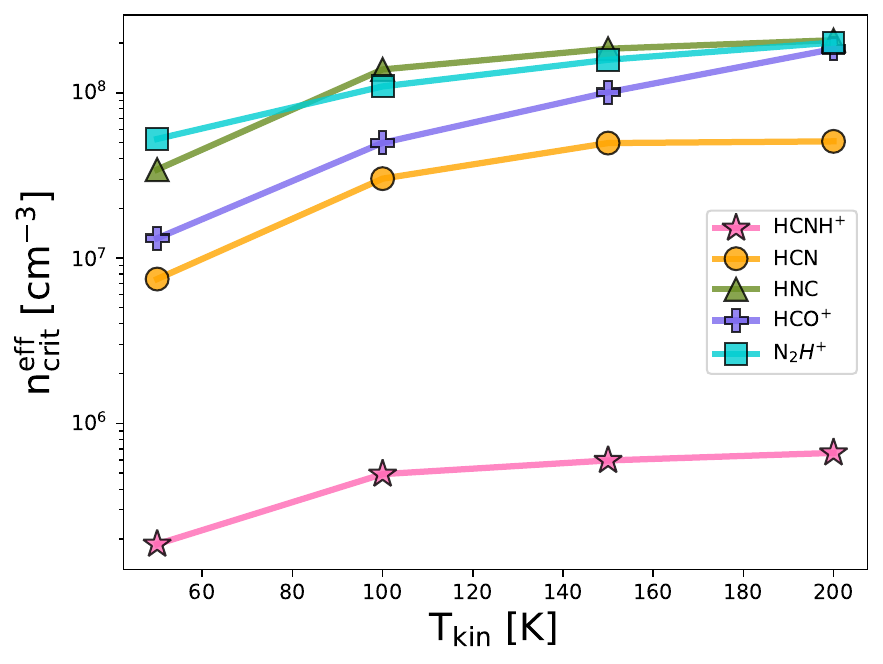}
\caption {Expected rotational temperature of non-LTE models and effective critical densities for the bands analysed in the work. In both panels the same ro-vibrational bands are shown: HCNH$^+$  ($\nu_5$; pink stars), HCN (orange circles), HNC (green triangles), HCO$^+$ (purple pluses), N$_2$H$^+$ (turquoise squares). Left panel: $T_{\rm rot}$ from simulated NLTE absorption bands as a function of the radiation temperature ($T_{\rm rad}$). The black dashed line indicate when  T$_{\text{rot}}$ is equal to T$_{\text{rad}}$.  The two HCN curves correspond to the two columns for bands 2$\nu_2$ (upper curve) and $\nu_2$ (bottom curve) reported in Table~\ref{tab:molecules}. Right panel: effective critical density of the ro-vibrational band vs. $T_{\rm kin}$.}
\label{fig:Trad}
\end{figure*}

 In this Section, we show Fig.~\ref{fig:Trad},  which is invoked to illustrate how the infrared radiative pumping mechanism can explain the range of $T_{\rm rot}$ values measured from the observed spectra. This figure is discussed in Section~\ref{lvg}, and in Table~\ref{tab:coeff} we report the relevant quantities mentioned in the discussion, namely: the Einstein A$_{ul}$ coefficients, the Einstein B$_{ul}$ coefficients normalized to that of HCO$^+$, the dipole moments, and $n_{\rm crit}^{\rm eff}$ at 200 K for each molecule presented in Fig.~\ref{fig:Trad}.

\section{Initial abundances of the chemical model}\label{apx:chemical}

\begin{figure*}
\centering
\includegraphics[width=1.\textwidth]{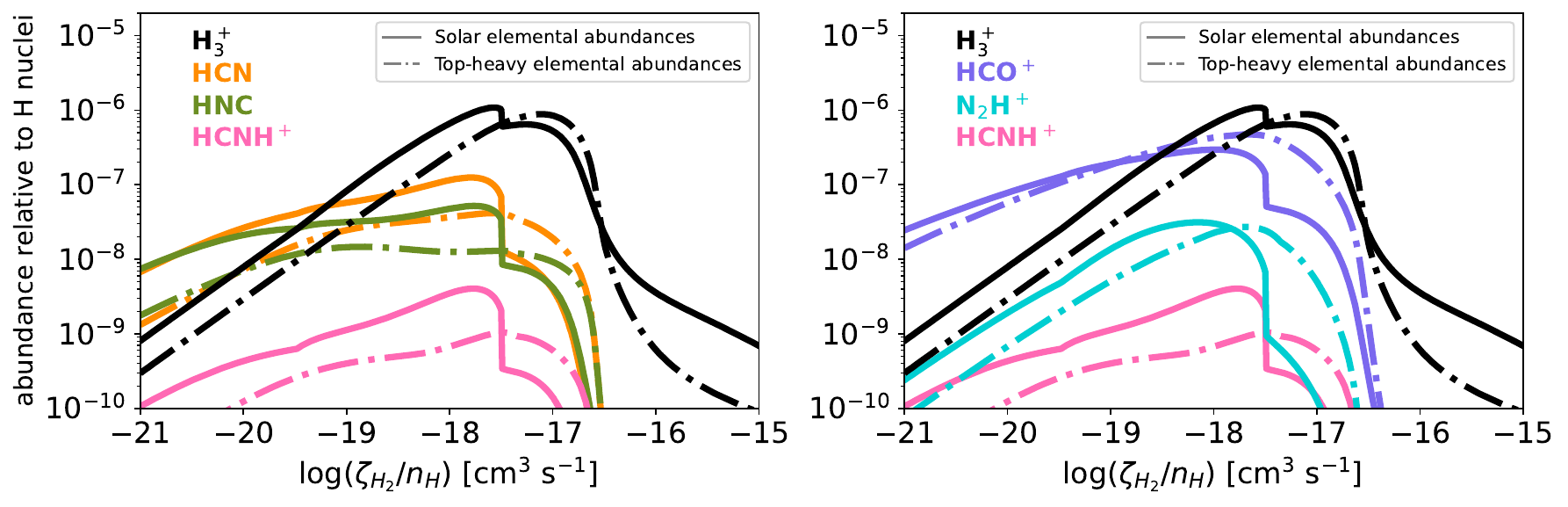}
\caption { Predictions of molecular fractional abundances as a function of the $\zeta_{H_2}/n_{\rm H}$ ratio. The solid curves are the same shown in Fig.~\ref{fig:models}, which are obtained using initial solar elemental abundances. The dash-dotted curves represent predictions for a cloud with top-heavy elemental abundances. The same colors as Fig.~\ref{fig:models} are used to distinguish the different molecules. }
\label{fig:top_heavy}
\end{figure*}

 We explored how predictions of molecular fractional abundances change when varying the initial elemental abundances in the chemical model. Roughly solar elemental abundances, with a C/O ratio of $\sim$ 0.54, are adopted in the models presented in Section~\ref{ss:chemical}. In Fig.~\ref{fig:top_heavy}, we compare these models (solid curves) with predictions using initial top-heavy elemental abundances (dash-dotted curves). The top-heavy abundances are obtained by increasing the oxygen abundance by a factor of four and the remaining elements by a factor two, resulting in a C/O ratio of $\sim$ 0.27,  typical of a top-heavy initial mass function (IMF; e.g., \citealt{Nomoto97}). Such initial conditions may be suitable in the case of recent star formation activity. The final predictions experience some changes if a top-heavy scenario is assumed. Essentially, CO increases its abundance while those molecules containing carbon (but not oxygen) decrease their abundance as a consequence of the lower C/O ratio. This affects to HCN and HNC. A further consequence is a general decrease in the abundance of cations such as N$_2$H$^+$ and HCNH$^+$, which is in part caused by an enhanced destruction of H$_3^+$ with CO. Although the abundances of HCN, HNC, and HCNH$^+$ decrease due to the lower availability of carbon in the top-heavy regime, the abundance ratio HCNH$^+$/(HCN +  HNC) remains relatively unchanged. It is also worth noting that the discontinuity due to bistability almost vanishes and the abundance peaks shift to higher $\zeta_{\rm H_2}/n_{\rm H}$ ratios.

\section{Chemical model compared to analytical predictions}\label{apx:comparison}

\begin{figure*}
\centering
\includegraphics[width=1.\textwidth]{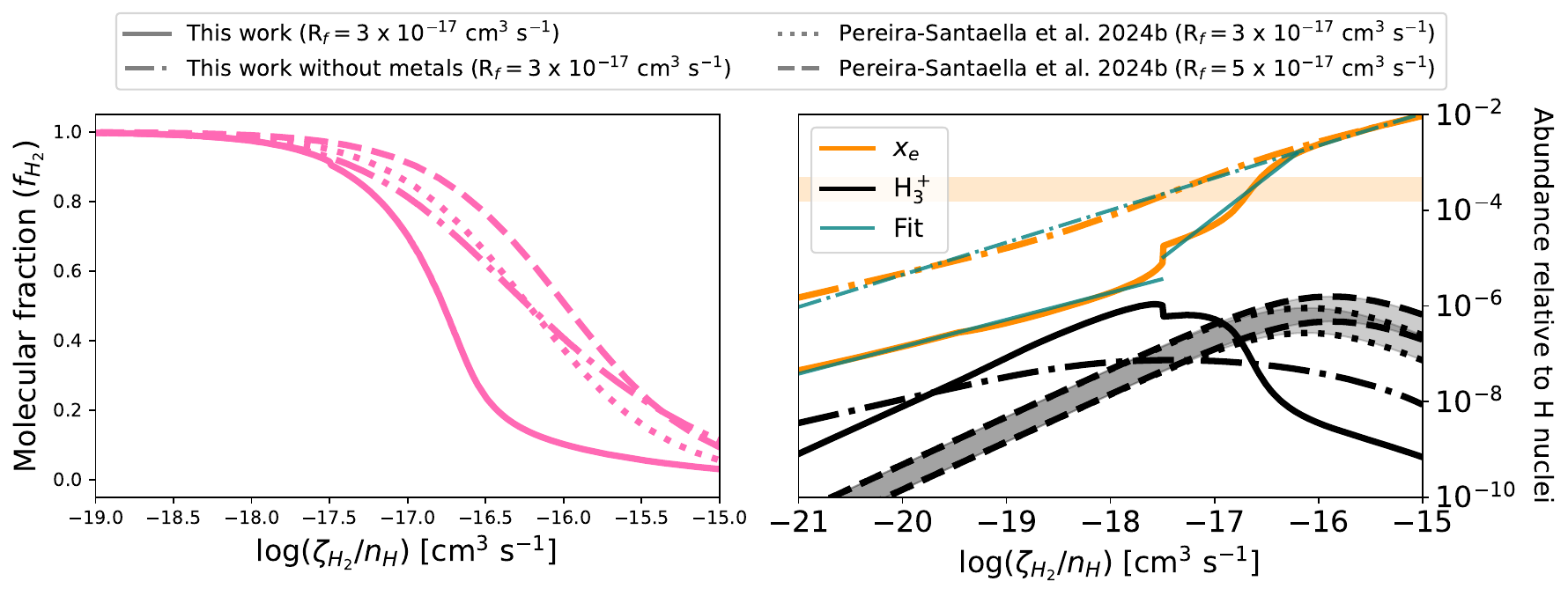}
\caption { Comparison between the molecular fraction, $f_{\rm H_2}$= 2$N({\rm H_2})/(2N({\rm H_2}) + N({\rm H}))$, (left panel) and the fractional abundances (right panel) predicted by our chemical model for obscured dense molecular clouds and those from \citet{Pereira24b} for diffuse and translucent clouds.
In both panels, the solid lines correspond to the model shown in Fig.~\ref{fig:models}, while the dash-dotted lines show the predictions obtained when metals are excluded from the chemical network. 
The dashed curve correspond to the predictions from \citet{Pereira24}, with R$_f$=5$\times$10$^{-17}$  cm$^3$ s$^{-1}$. The dotted lines represent analytical predictions using the same formulation but with R$_f$=3$\times$10$^{-17}$  cm$^3$ s$^{-1}$.  
In the right panel, electron and H$_3^+$ fractional abundances are shown in orange and black, respectively. The electron abundance, $x_e$, range adopted by \citet{Pereira24} is indicated by the shaded orange area. The green line is the best power-law fit to $x_e$ given by Eq.\,\ref{eq:fit1} and \ref{eq:fit2}.} \label{fig:hydro_frac}

\end{figure*}

\citet{Pereira24} predicted the H$_3^+$ fractional abundance as function of $\zeta_{\rm H_2}\slash n_{\rm H}$ using Eq.~2 of \citet{Neufeld17}, which was derived for diffuse and translucent clouds. Following \citet{Gonzalez13}, they adopted a constant electron fractional abundance in the range $x_e =$ (1.5–5) $\times \, 10^{-4}$  (derived from models by \citealt{Bruderer09} for relative high values of $\zeta_{\rm H_2}/n_{\rm H}$) and a molecular fraction ($f_{\rm H_2}$) given by the maximum value of Eq.~2 of \citet{Gonzalez13}. There are two main differences compared to the chemical model used in this work: (1) the analytical formulation assumes a constant electron abundance, and (2) metals are not included in the calculations (only $x_e$ and $f_{\rm H_2}$ are considered).  
In Fig.~\ref{fig:hydro_frac}, we compare these predictions.  
Our molecular fraction (solid pink line) starts decreasing at lower log($\zeta_{\rm H_2}/n_{\rm H}$) than their $f_{\rm H_2}$ (dashed pink line). This is due to the lower R$_f$ assumed in this work (3 $\times$ 10$^{-17}$ vs. 5 $\times$ 10$^{-17}$ cm$^{3}$ s$^{-1}$) and due to the significant destruction of H$_2$ by OH$^+$ and O$^+$ in the log($\zeta_{\rm H_2}/n_{\rm H}$ [cm$^{3}$ s$^{-1}$]) $\sim$ -17 to -15 range.
This difference is responsible for the shift of the H$_3^+$ abundance peak to lower $\zeta_{\rm H_2}/n_{\rm H}$ of 10$^{-17.6}$\,cm$^{3}$\,s$^{-1}$ versus the peak at $\zeta_{\rm H_2}/n_{\rm H}$=10$^{-15.9}$\, cm$^{3}$\,s$^{-1}$ predicted by the analytical model.

If metals are excluded from our chemical network, the predicted molecular fractions match the analytical results obtained under consistent assumptions (i.e., using R$_f$ = 3$\times$10$^{-17}$ cm$^3$ s$^{-1}$ instead of the 5$\times$10$^{-17}$  cm$^3$ s$^{-1}$; see dash-dotted and dotted pink curves). 
However, under these conditions, our model matches the analytical H$_3^+$ abundance only in the range $-$17.8 $\leq$ log($\zeta_{\rm H_2}/n_{\rm H}$) [cm$^3$ s${-1}$] $\leq$ $-$17.3, where the electron abundances from the chemical model overlap with the range assumed by \citealt{Pereira24}. 
Outside this interval, the electron abundance in our chemical model is greater at log($\zeta_{\rm H_2}/n_{\rm H}$) [cm$^3$ s${-1}$] $\geq$ $-$17.3 (lower at  log($\zeta_{\rm H_2}/n_{\rm H}$) [cm$^3$ s${-1}$] $\leq$ $-$17.8) than the constant range, thereby enhancing (reducing) H$_3^+$ destruction. This results in lower (higher) H$_3^+$ abundances (see the dash-dotted black line) compared to the predictions based on a constant $x{\rm_e}$. 

In the right panel of Fig.~\ref{fig:hydro_frac}, we present a power-law fit to the $x_e$ predictions given by the chemical model. For the model in which metals are not included we obtain
\begin{equation}
\log(x_e) = 0.67 \log(\zeta_{\rm H_2}/n_{\rm H}) + 8.14,
\label{eq:fit1}
\end{equation}
which corresponds to the green dash-dotted line.
In the case in which metals are included, we performed a piecewise power-law  fit of $x_e$ in two ranges in order to avoid the discontinuity at $\log \zeta_{\rm H_2}/n_{\rm H}$$\sim$$-$17.5.
\begin{small}
\begin{equation}
\log(x_e) = 
\begin{cases} 
    0.56 \log(\zeta_{\rm H_2}/n_{\rm H}) + 4.46  & \text{for } -21 < \log \zeta_{\rm H_2}/n_{\rm H} < -17.5 \\
    1.68 \log(\zeta_{\rm H_2}/n_{\rm H}) + 24.52  & \text{for } -17.5 < \log \zeta_{\rm H_2}/n_{\rm H} < -16.2.
\end{cases}
\label{eq:fit2}
\end{equation}
\end{small}
For higher $\zeta_{H_2}$/$n_H$, $x_e$ from the two models overlaps.

In conclusion, the differences between our predictions and those of \citet{Pereira24} arise from their model assumptions for diffuse and translucent clouds. Specifically, neglecting the role of metals and assuming a constant range of $x_e$ can result in higher $\zeta_{\rm H_2}/n_{\rm H}$ estimates.

\end{document}